\documentclass[twocolumn,showpacs]{revtex4}
\usepackage[dvips]{graphicx}
\usepackage{color}
\begin{document}
\title{How to make a traversable wormhole from a Schwarzschild black hole}
\author{Sean A. Hayward}
\email{sean_a_hayward@yahoo.co.uk} \affiliation{Department of Science
Education, Ewha Womans University, Seoul 120-750, Korea}
\author{Hiroko Koyama}
\email{koyama@gravity.phys.waseda.ac.jp} \affiliation{Department of Physics,
Waseda University, Okubo 3-4-1, Shinjuku-ku, Tokyo 169-8555, Japan}
\date{17th June 2004}

\begin{abstract}
The theoretical construction of a traversable wormhole from a Schwarzschild
black hole is described, using analytic solutions in Einstein gravity. The
matter model is pure phantom radiation (pure radiation with negative energy
density) and the idealization of impulsive radiation is employed.
\end{abstract}
 \maketitle

{\em Introduction.} The recently discovered acceleration of the universe
\cite{Spe,Kra} indicates that its large-scale evolution is dominated by unknown
dark energy which violates at least the strong energy condition, and perhaps
also the weak energy condition, where it is known as phantom energy \cite{Cal}.
Such phantom energy is precisely what is needed to support traversable
wormholes \cite{MT,Vis,HV1,HV2,wh,IH}. This Letter shows how one could,
strictly according to Einstein gravity, theoretically construct such a wormhole
from the most standard black-hole solution due to Schwarzschild.

\begin{figure}[t]
\includegraphics[height=28mm]{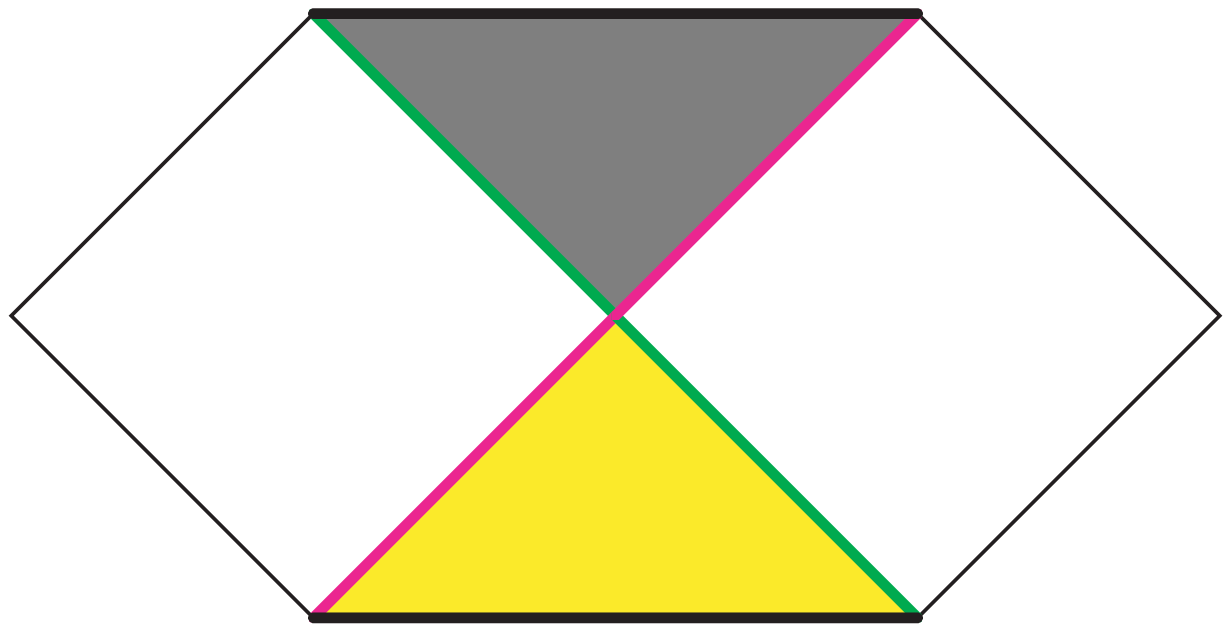}\hspace{2mm}
\includegraphics[height=28mm]{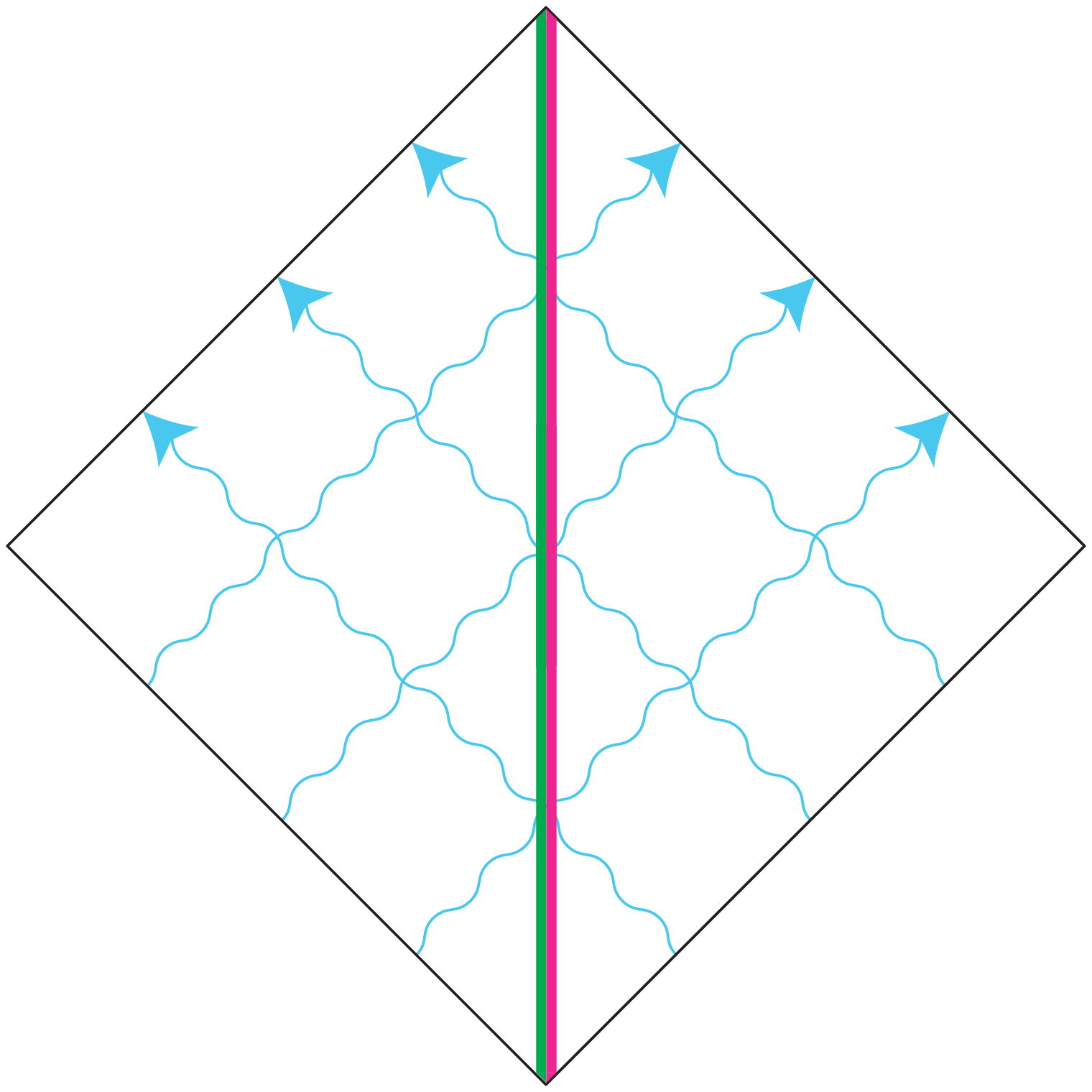}
\caption{Penrose diagrams of a Schwarzschild black hole and a traversable
wormhole \cite{pr}. The bold magenta and green lines represent the trapping
horizons, $\partial_+A=0$ and $\partial_-A=0$, respectively. They constitute
the Killing horizons of the black hole and the throat of the wormhole. Yellow
and gray quadrants represent past trapped and future trapped regions,
respectively. Wavy cyan lines represent the constant-profile phantom radiation
supporting the wormhole structure. }
 \label{fig:bh-wh}
\end{figure}

\begin{figure}[t]
\begin{center}
\includegraphics[width=7cm]{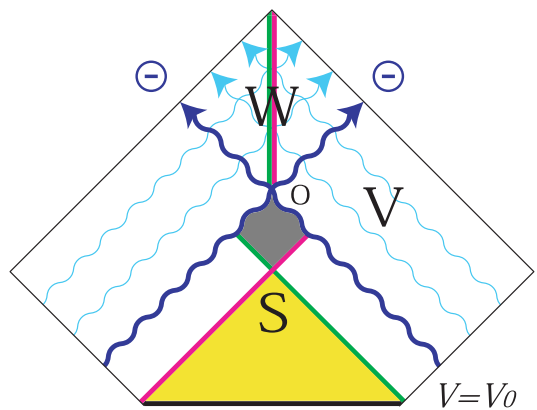}
\caption{Penrose diagram of the wormhole construction model. The wavy blue
lines represent impulsive radiation with negative energy density. The region S
is Schwarzschild, V is Vaidya and W is static-wormhole. } \label{fig:sudden}
\end{center}
\end{figure}

As the matter model, for simplicity we assume pure phantom radiation, i.e.\
pure radiation with negative energy density. It was recently shown that two
opposing streams of such radiation support a static traversable wormhole
\cite{pr}. Penrose diagrams of the wormhole space-time and the Schwarzschild
space-time are shown in Fig.\ref{fig:bh-wh}. The aim then is to find analytic
solutions to the field equations of General Relativity which interpolate
appropriately between such space-time regions. For simplicity again, we employ
the idealization of impulsive radiation, where the radiation forms an
infinitely thin null shell, thereby delivering finite energy-momentum in an
instant. It is then possible to explicitly construct spherically symmetric
solutions of the desired type using Vaidya regions, which describe space-times
with a single stream of pure radiation \cite{Vai}. The conformal diagram is
shown in Fig.\ref{fig:sudden}: one begins with a Schwarzschild region including
part of the black-hole region, then beams in impulses of phantom radiation from
both sides symmetrically. The impulses are followed by phantom radiation with
constant energy profiles, forming Vaidya regions. If the energies and timing
are related appropriately, the region left between the receding impulses after
collision is a static traversable wormhole. The analytic details of how to
match such regions using the Barrab\`es-Israel formalism \cite{BI} are quite
complex and given in a longer article \cite{KH}. However, it turns out that one
can understand the matching in a comparatively simple way, by continuity of the
area $A$ and a jump formula for the energy $E$ across impulses. This is the
method explained in this Letter.

Here $A$ is the area of the spheres of symmetry and $E$ is the active
gravitational mass-energy defined by \cite{MS,sph}
\begin{equation}
E=(1-g^{-1}(dr,dr))r/2
\end{equation}
where $r=\sqrt{A/4\pi}$ is the area radius and $g$ the space-time metric. Note
that $E=r/2$ on a trapping horizon $g^{-1}(dr,dr)=0$ \cite{bhd,1st}, which
includes both the Killing horizons of a stationary black hole, where $dr$ is
null, and the throat of a static wormhole, where $dr$ vanishes \cite{wh}. Thus
for a Schwarzschild black hole with mass $M$, $r=2M$ on the horizons, while
$r=a$ on a wormhole throat with area $4\pi a^2$.

{\em Basic solutions.} The required metrics are as follows. (i) The
Schwarzschild metric is given by
\begin{equation}
\label{metric:Schwarzschild}ds^2=r^2d\Omega^2
+\left(1-\frac{2M}r\right)^{-1}dr^2-\left(1-\frac{2M}r\right)dt^2
\end{equation}
where $d\Omega^2$ refers to the unit sphere and the constant $M$ is the
Schwarzschild mass, which coincides with the energy, $E=M$.

(ii) The metric of the Vaidya solutions is given by
\begin{equation}
\label{metric:Vaidya}ds^2=r^2d\Omega^2-dV\left[\left(1-\frac{2m}r\right)dV+2\zeta
dr\right]
\end{equation}
where $\zeta$ is a sign factor, with $\zeta =1$ for outgoing radiation and
$\zeta =-1$ for ingoing radiation. The mass function $m(V)$ coincides with the
energy, $E=m$. The corresponding energy tensor is
\begin{equation}\label{em:Vaidya}
T=-\frac\zeta{4\pi r^{2}}\frac{dm}{dV}dV\otimes dV.
\end{equation}
The Vaidya solutions reduce to the Schwarzschild solution in the case $m=M$
(constant), as can be seen by writing the latter in terms of the
Eddington-Finkelstein coordinate
\begin{equation}\label{EF}
V=t-\zeta\left(r+2M\ln(1-r/2M)\right)
\end{equation}
appropriate to the interior of the black hole, $r<2M$.

(iii) The static wormholes \cite{pr} supported by opposing streams of pure
phantom radiation can be written as
\begin{equation}\label{metric:wh}
ds^2=r^2d\Omega^2+2are^{-l^2}dl^2-\frac{2\lambda ae^{-l^2}}rdt^2
\end{equation}
where $t$ is a static time coordinate,
\begin{equation}
r(l)=a(e^{-l^2}+2l\phi),\qquad\phi(l)=\int_0^le^{-\ell^2}d\ell
\end{equation}
and $a>0$ and $\lambda>0$ are constants. The space-time is not asymptotically
flat, but otherwise constitutes a Morris-Thorne wormhole \cite{MT}, with doubly
minimal surfaces $dr=0$ at the throat $l=0$ and throat radius $r=a$. The energy
evaluates as $E=\epsilon$ where
\begin{equation}\label{eps}
\epsilon(l)=(e^{-l^2}+2l\phi-2e^{l^2}\phi^2)a/2.
\end{equation}
In terms of dual-null coordinates
\begin{equation}
x^\pm=t\pm{a\over{2\sqrt{\lambda}}}\left(le^{-l^2}+(1+2l^2)\phi\right)
\end{equation}
the metric is given by
\begin{equation}
ds^2=r^2d\Omega^2-\frac{2\lambda ae^{-l^2}}rdx^+dx^-.
\end{equation}
Then the energy tensor is found as
\begin{equation}
\label{em:wh} T=-\frac\lambda{8\pi r^2}(dx^+\otimes dx^++dx^-\otimes dx^-).
\end{equation}
This is the energy tensor of two opposing streams of pure phantom radiation,
with $\lambda=-4\pi r^2T_{tt}$ being the resulting negative linear energy
density. One may also write
\begin{equation}\label{metric:wh2}
ds^2=r^2d\Omega^2-\frac{\sqrt{\lambda}}{e^{l^2}\phi}dx^\pm
\left[\frac{2\sqrt{\lambda}a\phi}rdx^\pm\mp2dr\right]
\end{equation}
for comparison with the Vaidya solutions.

{\em Jump in energy due to impulsive radiation.} A general spherically
symmetric metric can be written in dual-null form as
\begin{equation}
 ds^2=r^2d\Omega^2-h\,dx^+dx^-
\end{equation}
where $r\ge0$ and $h>0$ are functions of the future-pointing null coordinates
$(x^+,x^-)$. Writing $\partial_\pm=\partial/\partial x^\pm$, the propagation
equations for the energy are obtained from the Einstein equations as \cite{1st}
\begin{equation}
\partial_\pm E=8\pi h^{-1}r^2(T_{+-}\partial_\pm r-T_{\pm\pm}\partial_\mp r).
\end{equation}
We consider impulsive radiation defined by
\begin{equation}
T=\frac{\mu_\pm dx^\pm\otimes dx^\pm}{4\pi r^2}\delta(x^\pm-x_0)
\end{equation}
where $\delta$ is the Dirac distribution, the constant $x_0$ gives the location
of the impulse and the constant $\mu_\pm$ is its energy. More invariantly, the
vector $\wp=-g^{-1}(\mu_\pm dx^\pm)$ is the energy-momentum of the impulse.
Then the jump
\begin{equation}
[E]_\pm=\lim_{\alpha\to0}\int_{x_0-\alpha}^{x_0+\alpha}\partial_\pm E\,dx^\pm
\end{equation}
in energy across the impulse is given by the jump formula
\begin{equation}
[E]_\pm=c^\pm\mu_\pm,\qquad c^\pm=-2(h^{-1}\partial_\mp r)\vert_{x^\pm=x_0}.
\end{equation}
The vector $c=c^+\partial_++c^-\partial_-$ is actually $c=g^{-1}(dr)$ and so
\begin{equation}
[E]_\pm=-\wp\cdot dr
\end{equation}
is a manifestly invariant form of the jump formula. Note that while the
energy-momentum vector $\wp$ (or $\mu_\pm dx^\pm$) is invariant, the energy
$\mu_\pm$ depends on the choice of null coordinate $x^\pm$, reflecting the fact
that a particle moving at light-speed has no rest frame and no preferred
energy. However, in a curved but stationary space-time, the stationary Killing
vector provides a preferred frame and a preferred energy $\mu_\pm$.

We need employ the jump formula only in the following cases. (i) Inside a
Schwarzschild black hole, we can take future-pointing $x^\pm=-r^{\ast}\pm t$
where $dr^{\ast}/dr=(2M/r-1)^{-1}$. Then $r^{\ast}=-(x^++x^-)/2$, $h=2M/r-1$
and $\partial_\pm r=(dr/dr^{\ast})\partial_\pm r^{\ast}=(1-2M/r)/2$ gives
$c^\pm=1$. (ii) Inside a Schwarzschild white hole, we can take future-pointing
$x^\pm=r^{\ast}\pm t$ and similarly obtain $c^\pm=-1$. (iii) On the throat of a
static wormhole, where $\partial_+r=\partial_-r=0$ and $h$ is finite \cite{MT},
one finds $c^\pm=0$. Then the jump formula yields
\begin{equation}
[E]_\pm=\left\{\begin{array}{ll} \mu&\quad\hbox{inside a
Schwarzschild black hole}\\
-\mu&\quad\hbox{inside a Schwarzschild white hole}\\
0&\quad\hbox{on the throat of a static wormhole}
\end{array}
\right.
\end{equation}
where the subscripts on $\mu_\pm$ are now omitted.

\begin{figure}[t]
\includegraphics[width=3cm]{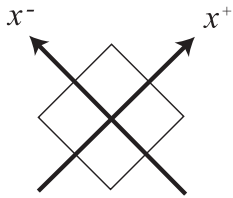}
\caption{An infinitesimal box around the intersection of two radiative
impulses.} \label{fig:intersect}
\end{figure}

{\em Wormhole construction.} The free data for the solution will be the initial
Schwarzschild mass $M>0$ and the energy $\mu<0$ of the impulses. Now imagining
an infinitesimal diamond-shaped box around the point O where the impulses
collide as in Fig.\ref{fig:intersect}, the energy $E$ will jump by $\mu$ from
the region S to V and by $0$ from the region V to W, evaluated in the limit at
the point. By continuity of the area $A=4\pi r^2$ and the fact that $E=a/2$ on
a wormhole throat of radius $a$, we therefore have $a/2=M+\mu$. Thus we require
$|\mu|<M$ and have determined $a$. The timing of the impulses is also
determined, as follows. The tortoise coordinate $r^{\ast}$ inside a
Schwarzschild black hole can be defined as
\begin{equation}
\label{tortoise} r^{\ast}=-r-2M\ln(1-r/2M).
\end{equation}
The symmetry of the impulses means that the intersection point O is given by
$t=0$, $r=a$ or $r^{\ast}=a^{\ast}$. Then the ingoing ($\zeta=-1$)
Eddington-Finkelstein relation (\ref{EF}) at the point O where the impulses
collide gives
\begin{equation}\label{timing}
V_0=-a^{\ast}=2(M+\mu)+2M\ln(-\mu/M).
\end{equation}
Finally, the mass function $m$ of the Vaidya regions is now implicitly
determined by comparing the expressions for the energy densities of the Vaidya
(\ref{em:Vaidya}) and static-wormhole (\ref{em:wh}) regions, or equivalently by
using the expression (\ref{eps}) for $E$ in the wormhole region and the jump
formula. Explicit expressions require a comparison of the null coordinates in
(\ref{metric:Vaidya}), (\ref{metric:wh2}) and are given in the longer article
\cite{KH}, which performs the matching in detail using the Barrab\`es-Israel
formalism \cite{BI}.

The results are analogous to those obtained for a two-dimensional gravity
theory where the solutions are much simpler \cite{HKL,KHK}. The opposite
process, namely collapse of a traversable wormhole to a black hole, is easy to
see for the wormhole considered here \cite{pr} and has previously been
demonstrated numerically for the Ellis wormhole \cite{Ell}, where still
unexplained critical phenomena were discovered \cite{SH}. Nonetheless, the
above solution appears to be the first which describes construction of a
traversable wormhole from a black hole in standard Einstein gravity.

\begin{figure}[t]
\begin{center}
\includegraphics[width=8cm]{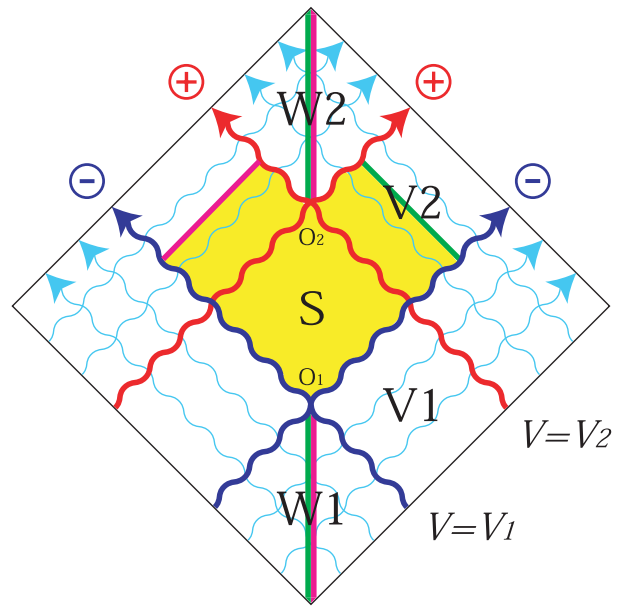}
\caption{Penrose diagram of the wormhole enlargement model. Wavy red lines
represent impulsive radiation with positive energy density. The regions W1 and
W2 are static-wormhole, V1 and V2 are Vaidya, and S is Schwarzschild.}
\label{fig:enlarge}
\end{center}
\end{figure}

{\em Wormhole enlargement.} As constructed above, the area of the wormhole
throat is less than the area of the initial black hole. Therefore another
interesting question is how to enlarge a wormhole. Similar analytic solutions
describing wormhole enlargement are given in the longer article \cite{KH} and
briefly described here. As in Fig.\ref{fig:enlarge}, one begins with a static
wormhole, beams in negative-energy impulses while simultaneously switching off
the constant-profile radiation, then beams in positive-energy impulses while
simultaneously switching on more constant-profile radiation. Then the regions
V1 and V2 are Vaidya, the region S is vacuum and therefore Schwarzschild, and
in fact is a white-hole region and therefore expanding.

The free data can be taken as the initial throat radius $a_1$ and the energies
$\mu_1<0$, $\mu_2>0$ of the impulses; this determines the energies of the
constant-profile radiation, the timings $V_1$, $V_2$ of the impulses, similarly
to (\ref{timing}), and the final throat radius $a_2$. In particular, the energy
$E$ will jump by $0$ from W1 to V1 and by $-\mu_1$ from V1 to S at the point
$\rm O_1$, and by $-\mu_2$ from S to V2 and by $0$ from V2 to W2 at $\rm O_2$,
yielding $M=a_1/2-\mu_1$ and $a_2/2=M-\mu_2$, where $M$ is the Schwarzschild
mass of S. Thus the increase in throat radius is given by
$a_2-a_1=-2(\mu_1+\mu_2)$ and so $|\mu_1|>\mu_2$.

The results are similar to those found in the two-dimensional model \cite{KHK}.
Self-inflating wormholes were also recently discovered numerically \cite{SH}
and wormhole inflation was previously suggested by Roman \cite{Rom}, but the
above solutions appear to be the first analytic examples of wormhole
enlargement in Einstein gravity with a specified exotic matter model.
Additionally, the enlargement is not a runaway inflation but an apparently
stable process, whereby the amount of enlargement can be controlled by the
energy or timing of the impulses. Reducing the wormhole size can similarly be
achieved by reversing the order of the positive-energy and negative-energy
impulses \cite{KH}.

{\em Remarks.} The solutions described here provide concrete manifestations of
ideas about traversable wormholes which were explained previously \cite{wh}
using general results and intuition concerning trapping horizons, particularly
how they develop under strengthening or weakening positive or negative energy
density. Some useful ideas or results were: a general definition of wormhole
mouth, unified with a general local definition of black hole \cite{bhd}; a
proof of the necessary violation of the null energy condition at a
non-degenerate wormhole mouth; an invariant measure of the radial curvature or
``flare-out'' at the mouth using a definition of surface gravity previously
proposed for dynamical black holes \cite{1st}; and analogues of basic laws of
black-hole dynamics \cite{bhd,bhd2} for wormholes. A newer observation is that
impulses of positive or negative energy respectively shift a trapping horizon
discontinuously to the past or future along the impulse. For brief conference
reviews see \cite{conf}.

In the wormhole-construction solution, the white-hole region of the space-time,
usually regarded as non-physical, is not essential to the argument; it could be
excised and replaced with a regular region of matter, though the subsequent
evolution of the matter would make analytic solutions more difficult. On the
other hand, the spatial topology of the maximally extended Schwarzschild
solution is relevant, since topology change is classically forbidden unless
causal loops exist \cite{Ger,Haw}. Black holes formed by gravitational collapse
of supernova remnants would presumably not have the appropriate topology.

For this issue one may turn to Wheeler's space-time foam picture \cite{Whe},
where by a reasonable application of quantum principles and the geometric
nature of General Relativity, Planck-sized virtual black holes are expected to
continually form and disappear. To quote Morris \& Thorne \cite{MT}: ``One
could {\em imagine} an exceedingly advanced civilization pulling a wormhole out
of this\dots space-time foam and enlarging it''. We now have exact solutions in
Einstein gravity describing how a Schwarzschild black hole, perhaps formed in
space-time foam, may be converted into a traversable wormhole and enlarged to
usable size. This might also be relevant in the quantum-gravity epoch presumed
to begin the universe, where primordial black holes and wormholes might both be
formed. It is tempting to speculate whether primordial wormholes might still
survive; clearly it would depend on the nature of the dark energy.

The fascinating potential of traversable wormholes, as short cuts across the
universe and even as time machines, has already passed into popular culture as
science fiction. As science, they now appear to be less speculative than much
of theoretical physics by recent standards. We have assumed only well-proven
Einstein gravity with an idealized model of phantom energy. If the mysterious
cosmological dark energy is or can be phantom in nature, one could argue that
traversable wormholes are as much a prediction of General Relativity as black
holes.

\acknowledgements H.K. is supported by a JSPS Research Fellowship for Young
Scientists.

\end{document}